\documentclass[11pt,a4paper,reqno]{amsart}
\usepackage{multirow}
\usepackage[centertags]{amsmath}
\usepackage{amsfonts}
\usepackage{amssymb}
\usepackage{amsthm}
\usepackage{newlfont}
\usepackage{a4}
\usepackage{enumerate}
\usepackage[pdftex]{graphicx,color}
\theoremstyle{definition}
\newtheorem{defn}{Definition}[section]

\newtheorem{tvr}[defn]{Proposition}

\theoremstyle{remark}

\usepackage{bbm}

\newlength{\defbaselineskip}
\newcommand{\setlinespacing}[1]%
           {\setlength{\baselineskip}{#1 \defbaselineskip}}

\renewcommand{\i}{\mathrm{i}}
\newcommand{\map}{\rightarrow}

\newcommand{\q}{\quad}

\renewcommand{\epsilon}{\varepsilon}

\newcommand{\la}{\lambda}

\renewcommand{\rho}{\varrho}
\renewcommand{\phi}{\varphi}

\newcommand{\R}{{\mathbb{R}}}

\newcommand{\Com}{{\mathbb C}}
\newcommand{\Z}{\mathbb{Z}}

\newcommand{\set}[2]{\left\{#1 \, |\, #2 \right\}}

\newcommand{\wt}{\widetilde}

\begin{document}

\title[3D alternating exponential functions ]
{Three variable exponential functions\\ of the alternating  group}
\author{Ji\v{r}\'{i} Hrivn\'{a}k$^{1,3}$}
\author{Ji\v{r}\'{i} Patera$^{1,2}$}
\author{Severin Po\v{s}ta$^{1,4}$}

\date{\today}

\begin{abstract}\
New class of special functions of three real variables, based on the alternating subgroup of the permutation group $S_3$, is studied. These functions are used for Fourier-like expansion of digital data given on lattice of any density and general position. Such functions have only trivial analogs in one and two variables; a connection to the $E-$functions of $C_3$ is shown. Continuous interpolation of the three dimensional data is studied and exemplified.
\end{abstract}\
\maketitle
\noindent
$^1$ Centre de recherches math\'ematiques,
         Universit\'e de Montr\'eal,
         C.~P.~6128 -- Centre ville,
         Montr\'eal, H3C\,3J7, Qu\'ebec, Canada;
         \email{patera@crm.umontreal.ca}\\
$^2$ MIND Research Institute,
         3631 S. Harbor Blvd., Suite 200,
Santa Ana, CA 92704,  USA
         \\         
$^3$ Department of physics,
Faculty of nuclear sciences and physical engineering, Czech
Technical University in Prague, B\v{r}ehov\'a~7, CZ-115 19 Prague, Czech
republic; jiri.hrivnak@fjfi.cvut.cz\\
$^4$ Department of mathematics,
Faculty of nuclear sciences and physical engineering, Czech
Technical University in Prague, Trojanova 13, CZ-120 00 Prague, Czech
republic; severin.posta@fjfi.cvut.cz

\section{Introduction}
Functions on Euclidean space $\R^n$ which are symmetric with respect to the permutation group $S_n$ are often dealt with in various branches of physics, namely in quantum theory or in theory of integrable systems. It is natural to consider restriction of the symmetry to the subgroup $A_n$ of $S_n$ consisting of transformations $w$ with $\det w=1$. Of some importance may also be the fact that $A_n\subset SO(n)$, while $S_n\nsubseteq SO(n)$.

Functions, considered in the paper, form a new family of special functions, called the alternating  exponential functions. They were introduced in \cite{KPaltE} for $A_n$, involving general number, $n\geq3$, of variables, but otherwise have not appeared in the literature. Here they depend on the smallest (nontrivial) number of real variables, namely three. In 2D the functions would become a product of two simple exponential functions each depending on one variable only. In 1D the analogous functions cannot be defined.  Through the inclusion $A_n\subset S_n$ they are closely related to symmetric and antisymmetric exponential functions, based on $S_n$, which are described in \cite{KPexp}. Through the isomorphism $S_n\equiv W(SU(n))$ they are also related to the $C-$ and $S-$functions \cite{KP-S,KP-C} based on the Weyl group $W(SU(n))$ of $SU(n)$. The relation is a consequence of the isomorphism $A_n\equiv W^e(SU(n))$, where $W^e$ is the even subgroup of $W$. Hence the functions of the paper could have been obtained, in principle, from the $E-$functions of $W^e(SU(n))$, see \cite{KP-E}.

As a motivation for studying the functions of this paper, one can single out several reasons.
(i) They are relatively simple when expressed as sums of common exponential functions of tree real variables \eqref{defE} measured in orthogonal directions of the 3-dimensional real Euclidean space $\R^3$.
(ii) Their continuous orthogonality, when integrated over the finite region $F(A_3^{\mathrm{aff}})\subset\R^3$, as well as their discrete orthogonality when their values are sampled on a fragment $L_{0,0,N,1}\subset F(A_3^{\mathrm{aff}})$ of the cubic lattice, make them into a curious alternative to the simple concatenation of three 1--dimensional transforms taken in the orthogonal directions. Comparison of the two transforms deserves further exploration.
(iii) The alternating symmetry should be advantageous in describing quantum systems possessing such a symmetry, as well as in some problems of quantum information theory.

In general, the $E-$functions are related to irreducible characters of compact simple Lie groups $G$ through the $C-$ and $S-$functions of $G$. The $C-$ and $S-$functions are defined by the summation of exponential functions over an orbit of the Weyl group $W(G)$ \cite{KP-S,KP-C}. The $E$-functions are defined by the summation of exponential functions over the  even subgroup $W^e(G)$ \cite{KP-E}. These functions were recently studied as special functions with many practically
useful properties. In Lie theory they are known from the Weyl formula
for the characters of irreducible representations of the
compact Lie group~$G$.
Note that the group $A_n$ is not a reflection group. Therefore it has no associated root system.

The alternating  exponential functions of three variables \eqref{defE} are most likely to be used in 3D applications due to their relative simplicity and to the fact that  their variables are given with respect to an orthonormal basis. Practical aspects of their applicability still need to be evaluated and compared with other systems of orthogonal functions. In the lowest case properties, we want to emphasize, are transparent and can be exploited to details not accessible to the analysis in general dimension. The simplicity of the functions offers an advantage that may also be decisive. Our goals are  to describe (i) discretization of the functions, (ii) expansions of digital data given on lattices of any density, and (iii) the interpolation of the data in `Fourier space', i.e. in the space of the coefficients of the expansions.

In Section~2 the functions are defined and their basic properties are shown. Their relation to $E-$functions of the Lie group $C_3$, their continuous and discrete orthogonalities and the corresponding Fourier expansions are given. In Section~3 the interpolation with alternating group functions is described and exemplified. Concluding remarks are contained in Section~4.

\section{Three dimensional alternating exponential functions}

\subsection{Definitions, symmetries and general properties}\

Three dimensional alternating exponential
functions $E_{(\lambda,\mu,\nu)}\colon \R^3\map \Com$ have
the following explicit form
\begin{align}\label{defE}
 E_{(\lambda,\mu,\nu)}(x,y,z)
     &=\frac{1}{2}\left|\begin{matrix}
     e^{2\pi i\lambda x}&e^{2\pi i\lambda y}&e^{2\pi i\lambda z} \\
     e^{2\pi i\mu x}&e^{2\pi i\mu y}&e^{2\pi i\mu z} \\
     e^{2\pi i\nu x}&e^{2\pi i\nu y}&e^{2\pi i\nu z}
     \end{matrix}\right|
     +\frac{1}{2}\left|\begin{matrix}
     e^{2\pi i\lambda x}&e^{2\pi i\lambda y}&e^{2\pi i\lambda z} \\
     e^{2\pi i\mu x}&e^{2\pi i\mu y}&e^{2\pi i\mu z} \\
     e^{2\pi i\nu x}&e^{2\pi i\nu y}&e^{2\pi i\nu z}
     \end{matrix}\right|^+\\
     &=e^{2\pi i(\lambda x+\mu  y+\nu z)}
     +e^{2\pi i(\lambda  z+\mu x+\nu y)} \
     +e^{2\pi i(\lambda  y+\mu z+\nu x)}\,,
     \quad x,y,z,\lambda,\mu,\nu\in\R\,,\notag
\end{align}
where the determinant with superscript $^+$ stands for {\it permanent} \cite{Henryk}, which is symmetric with respect to permutations of its rows and columns.

Note that in less than three variables, the definition \eqref{defE} leads to common functions. For two variables, we have $E_{\lambda,\mu}(x,y)$ as a product of two exponential functions, each depending on one of the variables. For one-dimensional case we obtain $E_\lambda(x)=e^{2\pi i\lambda x}$.

From  \eqref{defE} we immediately have symmetry of $E_{(\lambda,\mu,\nu)}(x,y,z)$ with respect to cyclic permutations of variables~$(x,y,z)$ and $(\la,\mu,\nu)$
\begin{gather}\label{expant}
E_{(\lambda,\mu,\nu)}(x,y,z)=E_{(\lambda,\mu,\nu)}(z,x,y)=  E_{(\lambda,\mu,\nu)}(y,z,x)
=E_{(\nu,\lambda,\mu)}(x,y,z)=  E_{(\mu,\nu,\lambda)}(x,y,z).
\end{gather}
Therefore, we consider only functions $E_{(\lambda,\mu,\nu)}$ with so called {\it semidominant} $(\lambda,\mu,\nu)$, that is triples $(\lambda,\mu,\nu)$ with $\la\geq\mu\geq \nu$ or $\mu>\la>\nu$.
The set of all semidominant triples is denoted by $D^e_+$.

\begin{figure}[htp]
\includegraphics[width=5cm]{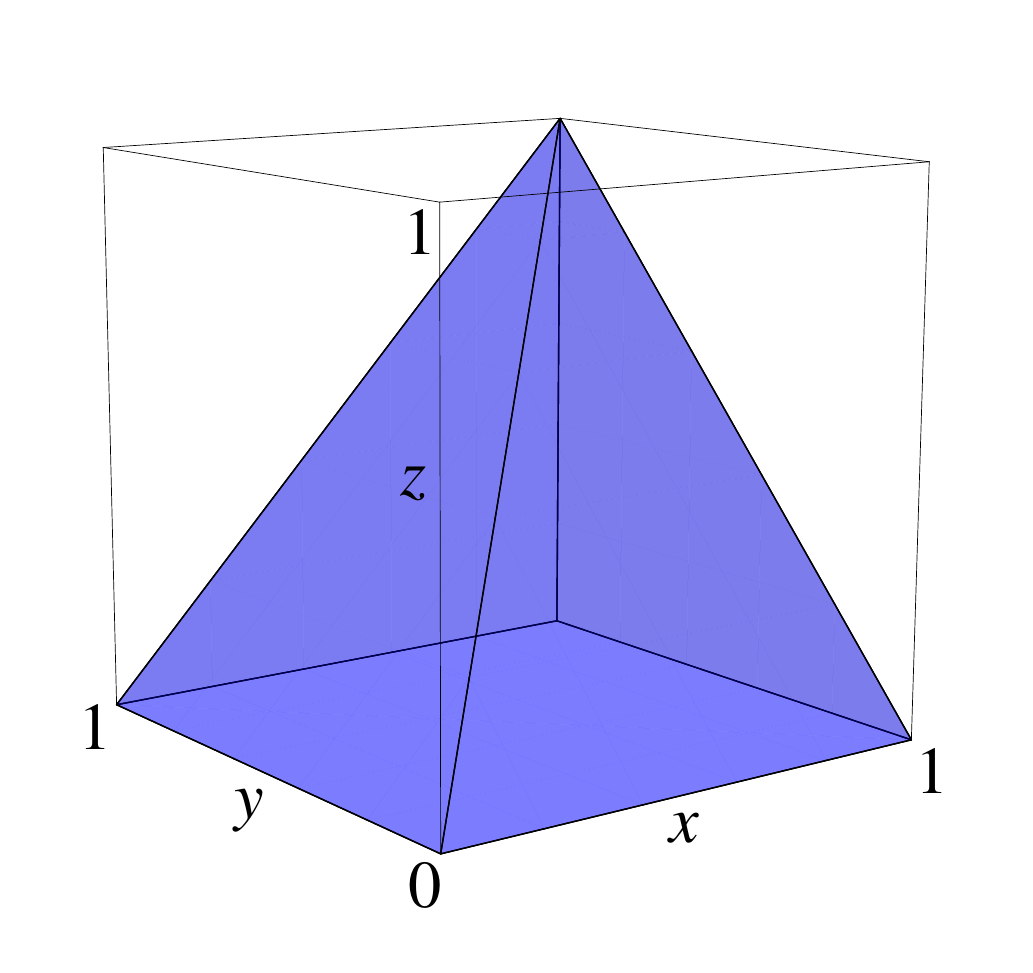}
\q\q\q
\includegraphics[width=5cm]{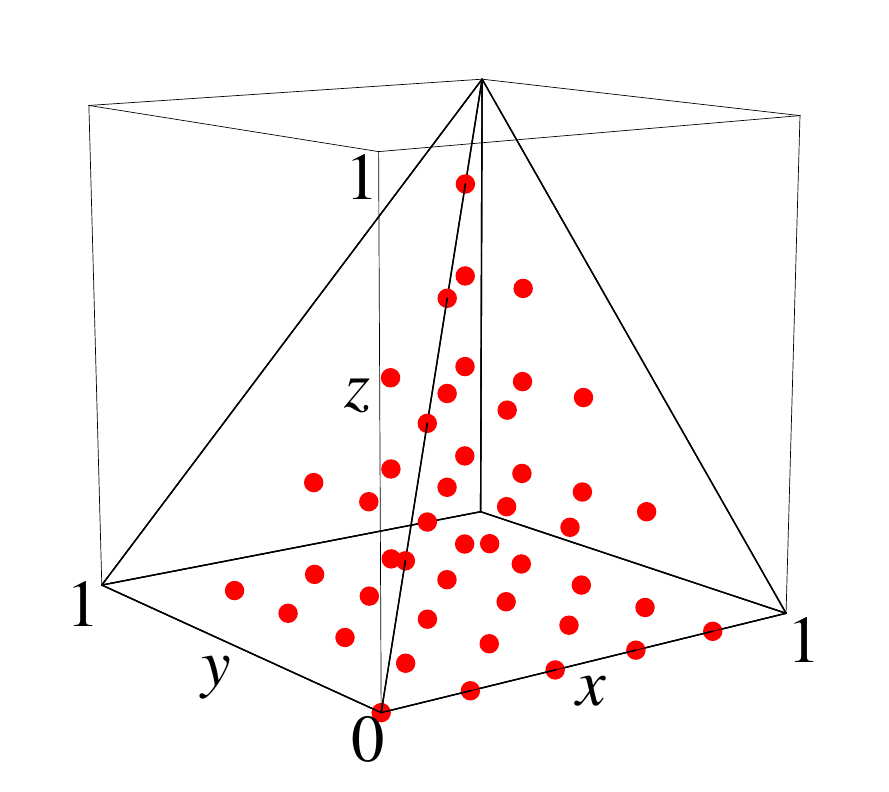}
\caption{The fundamental domain $F(A_3^{\mathrm{aff}})$ and the grid $L_{0,0,5,1}$, given by (\ref{origgrid}),
inside $F(A_3^{\mathrm{aff}})$.}\label{fig:funddom}
\end{figure}

The functions $E_{(k,l,m)}$ with $k,l,m\in\Z$ have additional symmetries induced by the periodicity of exponential function:
\begin{equation}\label{expper}
    E_{(k,l,m)}(x+r,y+s,z+t)= E_{(k,l,m)}(x,y,z),\q r,s,t\in \Z.
     \end{equation}
The relations (\ref{expant}) and (\ref{expper}) imply that it is sufficient to consider the functions $E_{(k,l,m)}$, $k,l,m\in\Z$  on the closure of the {\it fundamental domain} $F(A_3^{\mathrm{aff}})$~\cite{KPexp}. The fundamental domain $F(A_3^{\mathrm{aff}})$ can be chosen in 3D to be equal to the part of the cube shown on Figure~\ref{fig:funddom},
\begin{equation*}
F(A_3^{\mathrm{aff}})= \set{(x,y,z)\in(0,1)\times(0,1)\times(0,1)}{
x> z,\,y> z}.
\end{equation*}

For $a\in \R$, we also have
\begin{equation}\label{aexpshift}
 E_{(k,l,m)}(x+a,y+a,z+a)= e^{2\pi i(k+l+m)a}E_{(k,l,m)}(x,y,z).
\end{equation}

\subsection{Connection with $E-$orbit functions of the simple Lie group $C_3$}\

The $C-$ and $S-$functions of $C_3$ are defined by the summation of exponential functions over an orbit of the Weyl group $W(C_3)$. The $E-$functions are defined by the summation over the even subgroup $W^e(C_3)$.

Let us show that $E-$functions coincide, after a suitable linear substitution of variables, with the functions which are obtained by symmetrizing three dimensional alternating exponential functions $E_{(\lambda,\mu,\nu)}$ over the subgroup $W^e(C_3)$.
To make this connection clear we introduce the following non-orthogonal bases
expressed in terms of orthogonal vectors $e_1$, $e_2$, $e_3$:
\begin{equation*}
\begin{alignedat}{4}
\alpha_1&=e_1-e_2, &\qquad \omega_1&=e_1,         &\qquad \alpha^\vee_1&=2e_1-2e_2,  &\qquad  \omega^\vee_1&=2e_1,\\
\alpha_2&=e_2-e_3, &\qquad \omega_2&=e_1+e_2,     &\qquad \alpha^\vee_2&=2e_2-2e_3,  &\qquad  \omega^\vee_2&=2e_1+2e_2,\\
\alpha_3&=2e_3,    &\qquad \omega_3&=e_1+e_2+e_3, &\qquad \alpha^\vee_3&=2e_3,       &\qquad  \omega^\vee_3&=e_1+e_2+e_3.
\end{alignedat}
\end{equation*}

Assuming that $\alpha_1$ and $\alpha_2$ are the short simple roots
and $\alpha_3$ is the long simple root of $C_3$, we take $\langle e_i,e_j\rangle=\frac{1}{2}\delta_{ij}$, that is
$$
\langle\alpha_1,\alpha_1\rangle=1\,,\qquad
\langle\alpha_2,\alpha_2\rangle=1\,,\qquad
\langle\alpha_3,\alpha_3\rangle=2.
$$
The scalar product of two vectors, one given in
the $\omega$-basis and the other in the dual $\omega^\vee$-basis of $C_3$,
is calculated as follows:
\begin{align*}
\langle  v,\, \theta\rangle
&=
\langle (v_1 \omega_1+v_2 \omega_2 +v_3\omega_3),\,
(\theta_1 \omega^\vee_1+\theta_2 \omega^\vee_2+\theta_3 \omega^\vee_3)
\rangle
\\
&=(v_1+v_2+v_3)\theta_1+(v_1+2v_2+2v_3) \theta_2 +\left(
\frac{1}{2}v_1+v_2+\frac{3}{2}v_3\right)\theta_3.
\end{align*}

An orbit of $W(C_3)$ consists of all the distinct points (weights)
obtained from the dominant point ${v}$ by repeated
applications of reflections $r_{\alpha_1}$, $r_{\alpha_2}$  and $r_{\alpha_3}$
in mirrors orthogonal to the simple roots $\alpha_1$, $\alpha_2$, $\alpha_3$
of $C_3$, according to the formula
$$
r_{\alpha_j}{v}={v}-
\frac{2\langle\alpha_j,{v}\rangle}{\langle\alpha_j,\alpha_j\rangle}\alpha_j\,.
$$
A generic $W(C_3)$-orbit consists of 48 points/weights.
We consider only even subgroup $W^e(C_3)$ of the Weyl group which consists
of 24 elements. The generic orbit of the group $W^e(C_3)$ is written down
in Table \ref{tab:weightsC3}.

\begin{table}[ht]
\centering
\begin{tabular}{c l}
\hline\hline
\vrule height13pt depth 6pt width0pt
$l$ & weights \\
\hline
\vrule height14pt depth 0pt width0pt
$0$   & $(v_1,v_2,v_3)$\\
\vrule height14pt depth 0pt width0pt
$2$   & $(-v_1-v_2,v_1,v_2+v_3),(v_2,-v_1-v_2,v_1+v_2+v_3),(-v_1,v_1+v_2+2 v_3,-v_3),$\\
      & $(v_1+v_2+2 v_3,-v_2-2 v_3,v_2+v_3),(v_1+v_2,v_2+2 v_3,-v_2-v_3)$\\
\vrule height14pt depth 0pt width0pt
$4$   & $(-v_1-2 v_2-2 v_3,v_1+v_2,v_3),(-v_2-2 v_3,-v_1,v_1+v_2+v_3),$\\
      & $(-v_1-v_2-2 v_3,v_1+2 v_2+2 v_3,-v_2-v_3),(v_2+2 v_3,-v_1-2 v_2-2 v_3,v_1+v_2+v_3),$\\
      & $(-v_2,v_1+2 v_2+2 v_3,-v_1-v_2-v_3),(v_1+2 v_2+2 v_3,-v_1-v_2-2 v_3,v_3),$\\
      & $(v_2+2 v_3,v_1+v_2,-v_1-v_2-v_3),(v_1+2 v_2+2 v_3,-v_2,-v_3)$\\
\vrule height14pt depth 0pt width0pt
$6$   & $(-v_2,-v_1-v_2-2 v_3,v_1+v_2+v_3),(-v_2-2 v_3,v_1+v_2+2 v_3,-v_1-v_2-v_3),$\\
      & $(-v_1-v_2-2 v_3,-v_2,v_2+v_3),(-v_1-2 v_2-2 v_3,v_2+2 v_3,-v_3),$\\
      & $(v_2,v_1,-v_1-v_2-v_3),(v_1+v_2,-v_1-2 v_2-2 v_3,v_2+v_3),$\\
      & $(v_1+v_2+2 v_3,-v_1,-v_2-v_3)$\\
\vrule height14pt depth 10pt width0pt
$8$   & $(v_1,-v_1-v_2,-v_3),(-v_1,-v_2-2 v_3,v_3),(-v_1-v_2,v_2,-v_2-v_3)$\\
\hline \\
\end{tabular}
\caption{The generic orbit in the $\omega$-basis of the even subgroup of the Weyl group $W(C_3)$; weights are sorted
according to a minimal number $l$ of reflections used to generate them.}
\label{tab:weightsC3}
\end{table}

The generic $E-$function of $C_3$, denoted here by $E^W_{v}( \theta)$,
is defined in \cite{KP-E} as
\begin{equation*}
E^W_{ v}(\theta)
    =\sum_{w\in W^e(C_3)}e^{2\pi i\langle w v,\theta\rangle}.
\end{equation*}
After the following change of variables,
\begin{alignat*}{2}
\label{changeofvars}
v_1&=\lambda-\mu,   &\qquad  \theta_1&=x-y,\nonumber\\
v_2&=\mu-\nu,       &\qquad  \theta_2&=y-z,\\
v_3&=\nu,           &\qquad  \theta_3&=2z,\nonumber
\end{alignat*}
function $E^W$ takes the form
\begin{align*}
E^W_{ (v_1,v_2,v_3)}(\theta_1,\theta_2,\theta_3)={}&
e^{2i\pi  (-x \nu -y \mu -z \lambda )}+e^{2i\pi  (-x \nu +y \mu +z \lambda )}+e^{2i\pi  (-x \nu +y \lambda -z \mu )}+
e^{2i\pi  (-x \nu -y \lambda +z \mu )}\\+&
e^{2i\pi  (x \nu -y \mu +z \lambda )}+e^{2i\pi  (x \nu +y \mu -z
   \lambda )}+
   e^{2i\pi  (x \nu -y \lambda -z \mu )}+e^{2i\pi  (x \nu +y \lambda +z \mu )}\\+&
   e^{2i\pi  (-x \mu -y \nu +z \lambda )}+
e^{2i\pi  (x \mu -y \nu -z \lambda )}+e^{2i\pi  (-x \lambda -y \nu -z \mu )}+e^{2i\pi  (x
   \lambda -y \nu +z \mu )}\\+
   &e^{2i\pi  (-x \mu +y \nu -z \lambda )}+e^{2i\pi  (x \mu +y \nu +z \lambda )}+e^{2i\pi  (x \lambda +y \nu -z \mu )}+
   e^{2i\pi  (-x \lambda +y \nu +z \mu )}\\+&
   e^{2i\pi  (-x \mu -y \lambda -z \nu
   )}+e^{2i\pi  (x \mu +y \lambda -z \nu )}+
   e^{2i\pi  (x \lambda -y \mu -z \nu )}+e^{2i\pi  (-x \lambda +y \mu -z \nu )}\\+
   &e^{2i\pi  (-x \mu +y \lambda +z \nu )}+
   e^{2i\pi  (x \mu -y \lambda +z \nu )}+e^{2i\pi  (-x \lambda -y
   \mu +z \nu )}+e^{2i\pi  (x \lambda +y \mu +z \nu )}.
\end{align*}

As a group, $W^e(C_3)$ can be expressed as product $\widetilde W^e A_3$, where
$\widetilde W^e$ is a subgroup of order 8
generated by two elements,
\begin{equation*}
\widetilde W^e=\langle\lbrace r_{\alpha_1} r_{\alpha_3},
(r_{\alpha_2} r_{\alpha_3})^2 \rbrace \rangle,
\end{equation*}
and $A_3$ is subgroup of order 3
generated by rotation $r_{\alpha_1} r_{\alpha_2}$.
Then we have
\begin{equation*}
E^W_{ (\lambda-\mu,\mu-\nu,\nu)}(x-y,y-z,2z)=
\sum_{w \in \widetilde W^e} E_{(\lambda,\mu,\nu)}{(w(x,y,z))}.
\end{equation*}

\subsection{Product decomposition}\

The product of two three dimensional alternating functions evaluated at the same point $(x,y,z)$
can be easily decomposed to the sum of
alternating functions with suitable indices. Such product-to-sum decomposition has the following explicit form
\begin{align*}
E_{(\lambda,\mu,\nu)}(x,y,z)
E_{(\lambda',\mu',\nu')}(x,y,z)
={}&E_{(\lambda+ \lambda',\mu+ \mu',\nu+\nu')}(x,y,z)
+E_{(\lambda+\mu',\mu+ \nu',\nu+\lambda')}(x,y,z)\\
+&E_{(\lambda+ \nu',\mu+ \lambda',\nu+\mu')}(x,y,z).
\end{align*}

Analogously, we obtain a product-to-sum decomposition formula for one function $E_{(\la,\mu,\nu)}$ evaluated at two different points

\begin{align*}
E_{(\lambda,\mu,\nu)}(x,y,z)
E_{(\lambda,\mu,\nu)}(x',y',z')
={}&E_{(\lambda,\mu,\nu)}(x+x',y+y',z+z')
+E_{(\lambda,\mu,\nu)}(x+y',y+z',z+x')\\
+&E_{(\lambda,\mu,\nu)}(x+z',y+x',z+y').
\end{align*}

\subsection{Continuous orthogonality}\

The functions $E_{(k,l,m)}$ are mutually orthogonal on the
fundamental domain $F(A_3^{\mathrm{aff}})$, i.e. for any two semidominant triples $(k,l,m)$, $(k',l',m')\in D_e^+\cap\Z^3$ it holds
\begin{equation*}
    \int_{F(A_3^{\mathrm{aff}})} E_{(k,l,m)}(x,y,z)
\overline{E_{(k',l',m')}(x,y,z)}\,dx \,dy \,dz=G_{klm}
\delta_{kk'}\delta_{ll'}\delta_{mm'},  \end{equation*}
where the overline means complex conjugation and the symbol $G_{klm}$ is defined by
\begin{equation*}
G_{klm}=\begin{cases} 3 & \text{if $k=l=m$}, \\ 1 & \text{otherwise}. \end{cases}
\end{equation*}

Every function $f\colon\ \R^3\map\Com$, which satisfies $f(x,y,z)=f(z,x,y)=f(y,z,x)$, and is periodic, i. e. $f(x+r,y+s,z+t)= f(x,y,z)$, $r,s,t\in \Z$, and has
continuous derivatives, can be expanded in a series of alternating exponential
functions $E_{(k,l,m)}$:
\begin{align*}
f(x,y,z)={}&\sum_{(k,l,m)\in D_e^+\cap\Z^3} {\wt c}_{klm} E_{(k,l,m)}(x,y,z),\\
{\wt c}_{klm} ={}& G_{klm}^{-1}\int_{F(A^{\mathrm{aff}}_3)} f(x,y,z)
\overline{E_{(k,l,m)}(x,y,z)}\,dx\, dy\,dz.
 \end{align*}

\subsection{Eigenfunctions of the Laplace and related operators}\

The functions $E_{(\lambda,\mu,\nu)}$ are eigenfunctions of the Laplace
operator, which in the Cartesian coordinates takes the form
$\Delta=\partial_x^2+\partial_y^2+\partial_z^2$, so that we have
\begin{equation}
\label{Alaplas}
\Delta E_{(\lambda,\mu,\nu)}(x,y,z)=-4\pi^2 (\lambda^2+\mu^2+\nu^2)
E_{(\lambda,\mu,\nu)}(x,y,z).
\end{equation}
Laplace equation (\ref{Alaplas}) can be generalized. Let
$\sigma_k(y_1,y_2,y_3)$ be the $k$th elementary symmetric polynomial of
degree $k$. In particular,
\begin{equation*}
\sigma_1(y_1,y_2,y_3)=y_1+y_2+y_3,\q
\sigma_2(y_1,y_2,y_3)=y_1 y_2+y_1 y_3+y_2 y_3,\q
\sigma_3(y_1, y_2, y_3)=y_1 y_2 y_3.
\end{equation*}
Then we have
\begin{equation}
\label{Alaplas2}
\sigma_k(\partial_x^2,\partial_y^2,\partial_z^2) E_{(\lambda,\mu,\nu)}(x,y,z)
=(-4\pi^2)^k \sigma_k(\lambda^2,\mu^2,\nu^2)E_{(\lambda,\mu,\nu)}(x,y,z).
\end{equation}
The equations (\ref{Alaplas2}) are algebraically independent for $k=1,2,3$.

\section{Discrete orthogonality and alternating interpolation}

\subsection{Discrete orthogonality}\

The purpose of this section is to describe in detail discrete orthogonality of alternating functions and apply it to the interpolation problem of arbitrary functions defined in $F(A_3^{\mathrm{aff}})$. We follow notions known in standard Fourier analysis (see e.g. \cite{Stoer}) and generalize them to the set of alternating functions.

For any positive integer $N$ we consider a grid of the form
\begin{equation}\label{origgrid}
    L_{0,0,N,1}=\left\{\left(\frac{r}{N},\frac{s}{N},\frac{t}{N}  \right)\Bigm|
    (r,s,t) \in D^e_+(0,N-1)\right\},
\end{equation}
where
\begin{equation*}
    D^e_+(N_1,N_2)=\left\{(r,s,t)\,|\,r\geq s \geq t \text{ or } s>r>t ;\ r,s,t=N_1,N_1+1,N_1+2,...,N_2\right\}.
\end{equation*}
Discrete orthogonality of alternating exponential functions over this grid
was proved in~\cite{KPexp}. The positive integer $N$ fixes the density
of the grid inside $\overline{F(A_3^{\mathrm{aff}})}$, the grid contains
$\frac{1}{3}N(N^2+2)$ points.
For example, for $N=3$ the
following points are placed inside the grid:
\begin{eqnarray*}
L_{0,0,3,1}&=&\{(0, 0, 0), (\tfrac{1}{3}, 0, 0),
(\tfrac{1}{3}, \tfrac{1}{3}, 0), (\tfrac{1}{3}, \tfrac{1}{3},
\tfrac{1}{3}), (\tfrac{1}{3}, \tfrac{2}{3}, 0),
(\tfrac{2}{3}, 0, 0),\\
&& (\tfrac{2}{3}, \tfrac{1}{3}, 0),
(\tfrac{2}{3}, \tfrac{1}{3}, \tfrac{1}{3}), (\tfrac{2}{3}, \tfrac{2}{3},
0), (\tfrac{2}{3}, \tfrac{2}{3}, \tfrac{1}{3}), (\tfrac{2}{3},
\tfrac{2}{3}, \tfrac{2}{3})\}.
\end{eqnarray*}
A visual example of such grid (for $N=5$)
is shown in Figure~\ref{fig:funddom}.
For applications it may be convenient to consider the orthogonality over
more general type of grid. Besides the parameter $N$, we parametrize
the grid by numbers $a\in \R$ and $b\in [0,1]$. The
equidistant grid $L_{a,b,N,1}$ is given by
\begin{equation*}
L_{a,b,N,1}=\left\{(x_r,y_s,z_t)\,|\,(r,s,t) \in D^e_+(0,N-1)
\right \},
\end{equation*}
where
\begin{equation*}
    (x_r,y_s,z_t)=\left( a+\frac{r+b}{N},a+\frac{s+b}{N},a+\frac{t+b}{N}  \right).
\end{equation*}
Using the property (\ref{aexpshift}), we observe that the orthogonality relations from~\cite{KPexp} also hold over the grid $L_{a,b,N,1}$:
\begin{equation}\label{sdortho}
\sum_{(r,s,t)\in D^e_+(0,N-1)} G_{rst}^{-1}E_{(k,l,m)}(x_r,y_s,z_t)\overline{E_{(k',l',m')}(x_r,y_s,z_t)}=G_{klm} N^3\delta_{kk'}\delta_{ll'}\delta_{mm'},
\end{equation}
where $(k,l,m),\, (k',l',m')\in D^e_+(0,N-1)$.

\subsection{Alternating discrete Fourier transform}\

Suppose we have a discrete function $f\colon L_{a,b,N,1}\map \Com $ defined on the grid $L_{a,b,N,1}$. The {\it alternating discrete Fourier transform} of $f$ over $L_{a,b,N,1}$ is given by
\begin{equation}\label{Abetas}
\beta_{klm}=\frac{1}{G_{klm}N^3}\sum_{(r,s,t)\in D^e_+(0,N-1)}
G_{rst}^{-1} f(x_r,y_s,z_t)\overline{E_{(k,l,m)}(x_r,y_s,z_t)},
\end{equation}
where  $(k,l,m) \in D^e_+(0,N-1)$.
Orthogonality relation (\ref{sdortho}) immediately gives the inverse transform of the coefficients $\beta_{klm}$:
\begin{equation*}
    f(x_r,y_s,z_t)=\sum_{(k,l,m)\in D^e_+(0,N-1)}\beta_{klm}E_{(k,l,m)}(x_r,y_s,z_t).
\end{equation*}

\subsection{General 3D trigonometric interpolation}\

Let us consider a symmetrically placed cube in $\R^3$ with the length of its side $T\in \R$. For  $a\in \R$ the cube is given by
\begin{equation*}
K_{[a,a']}=[a,a']\times [a,a']\times [a,a'].
\end{equation*}
Let us also choose an arbitrary natural number $N$ and parameter $b\in [0,1]$
and consider a symmetrically
placed $N^3$-point
grid $\widetilde L_{a,b,N,T}=\left\{(x_r,y_s,z_t)\,|\,r,s,t=0\dots N-1
\right \}\subset K_{[a,a+T]} $
where
\begin{equation*}
  (x_r,y_s,z_t)=\left( a+\frac{r+b}{N}T,a+\frac{s+b}{N}T,a+\frac{t+b}{N}T  \right).
\end{equation*}
Suppose we have a given function $f\colon K_{[a,a+T]}\map \Com$ and a set of points $\widetilde L_{a,b,N,T}\subset K_{[a,a+T]}$. In the following we assume that $N$ is odd, $N=2M+1$. The three--dimensional {\it (trigonometric) interpolation problem} can be formulated in the following way: find a trigonometric interpolating polynomial of the form 
\begin{equation*}
    \psi_{N,T}(x,y,z)=\sum_{k,l,m=-M}^{M}c_{klm} e^{2\pi\i k \frac{x}{T}}e^{2\pi\i l\frac{y}{T}}e^{2\pi\i m\frac{z}{T}}
\end{equation*}
such that it coincides with $f$ on the grid $\widetilde L_{a,b,N,T}$, that means it satisfies for all $(x_r,y_s,z_t)\in \widetilde L_{a,b,N,T}$ the condition $\psi_{N,T}(x_r,y_s,z_t)=f(x_r,y_s,z_t)$. We set for simplicity $T=1$ and  we have the trigonometric interpolating polynomial $\psi_{N}\equiv\psi_{N,1}$ of the form
\begin{equation*}
 \psi_N(x,y,z)=\sum_{k,l,m=-M}^{M}c_{klm}e^{2\pi\i k x}e^{2\pi\i ly} e^{2\pi\i mz}
\end{equation*}
satisfying on the $K_{[a,a+1]}$
\begin{equation}\label{trig2}
    \psi_{N}(x_r,y_s,z_t)=f(x_r,y_s,z_t), \q r,s,t=0\dots N-1.
\end{equation}
Note that in all of the following formulas one can always recover
an arbitrary size $T$ by linear
transformation
\begin{equation*}
(x,y,z)\map \left(\frac{x}{T},\frac{y}{T},\frac{z}{T}\right).
\end{equation*}
For $N=2M+1$, the trigonometric interpolating polynomial $\psi_{N}$ has $N^3$ unknown coefficients $c_{klm}$ corresponding to $N^3$ constraints (\ref{trig2}).

Due to the orthogonality of exponential functions, the solution of the interpolation problem always exists, is unique and the coefficients $c_{klm}$ are
given by
\begin{equation*}
c_{klm}=\frac{1}{N^3}\sum_{r,s,t=0}^{N-1}f(x_r,y_s,z_t)e^{-2\pi\i k x_r}e^{-2\pi\i ly_s}e^{-2\pi\i lz_t}.
\end{equation*}

\subsection{Alternating interpolation}\

For interpolation with alternating exponential functions we consider the domain $K_{[a,a+1]}^A$
which we define as a shifted closure of the domain $F(A_3^{\mathrm{aff}})$:
\begin{equation*}\label{Afunds}
K_{[a,a+1]}^A=(a,a,a)+\overline{F(A_3^{\mathrm{aff}})}.
\end{equation*}
Note that $K_{[a,a+1]}^A$ is a part of the cube $K_{[a,a+1]}$.

For a given function $f\colon K_{[a,a+1]}^A\map \Com$ and a set of points $L_{a,b,N,1}\subset K_{[a,a+1]}^A$ we define an {\it alternating interpolating function}

\begin{equation}\label{Atrig}
    \psi^A_{N}(x,y,z)=\sum_{(k,l,m)\in D^e_+(-M,M)}c^A_{klm}E_{(k,l,m)}(x,y,z)
\end{equation}
satisfying
\begin{equation}\label{Atrig2}
    \psi^A_{N}(x_r,y_s,z_t)=f(x_r,y_s,z_t), \q (r,s,t) \in D^e_+(0,N-1).
\end{equation}

For $N=2M+1$, the alternating interpolating function $\psi^A_{N}$ has $\frac{1}{3} (2M+1)(4M^2+4M+3)$
unknown coefficients $c^A_{klm}$ given by the same number of constraints (\ref{Atrig2}).
Again due to the orthogonality of the alternating exponential functions we have the following:
\begin{tvr}\label{Aip}
There exists a unique alternating interpolating function (\ref{Atrig})
satisfying (\ref{Atrig2}). The coefficients $c^A_{klm}$ are
given for $N=2M+1$ by
\begin{equation}\label{Ainterodd}
c^A_{klm}=\frac{1}{G_{klm}N^3}\sum_{(r,s,t)\in D^e_+(0,N-1)}
G_{rst}^{-1}f(x_r,y_s,z_t)\overline{E_{(k,l,m)}(x_r,y_s,z_t)}.
\end{equation}
\end{tvr}

Instead of the direct calculation of the coefficients $c^A_{klm}$,
one can use the alternating discrete
Fourier transform (\ref{Abetas}), and the resulting
coefficients $\beta_{klm}$ transform to $c^A_{klm}$'s. By direct
comparison of (\ref{Ainterodd}) to (\ref{Abetas}), we obtain for $N=2M+1$
\begin{equation*}
\begin{split}
c^A_{klm}=\beta_{klm},\q & 0 \leq k,l,m \leq M,\\
c^A_{klm}=\beta_{l, m + 2 M + 1, k},\q &    0 \leq k,l \leq M,\ k < l,\ -M \leq m\leq -1,\\
c^A_{klm}=\beta_{m + 2 M + 1, k, l},\q & 0 \leq k,l\leq M,\ k \geq l,\ -M \leq m \leq -1,\\
c^A_{klm}=\beta_{l + 2 M + 1, m + 2 M + 1, k},\q & 0 \leq k \leq M,\ -M \leq l \leq -1,\\
c^A_{klm}=\beta_{k + 2 M + 1, l + 2 M + 1,m + 2 M + 1},\q & -M \leq k,l \leq -1,\\
c^A_{klm}=\beta_{m + 2 M + 1, k + 2 M + 1, l},\q & -M \leq k \leq -1,\  0 \leq l \leq M.
\end{split}
\end{equation*}

\subsection{Example of alternating interpolation}\

As a model function, we take the following smooth characteristic function:
\begin{equation}
\label{bumpfunc}
f_{\alpha,\beta,(x_0,y_0,z_0)}(x,y,z)=
\begin{cases} 1 & \text{if $r<\alpha$}, \\
0 & \text{if $r>\beta$},\\
e\, \exp \left(
\left(
\dfrac{r - \alpha}{\beta - \alpha}
\right)^2 - 1
\right)^{-1} & \text{otherwise},
\end{cases}
\end{equation}
where $r=\sqrt{(x-x_0)^2+(y-y_0)^2+(z-z_0)^2}$.

\begin{figure}[htp]
\includegraphics[width=7cm]{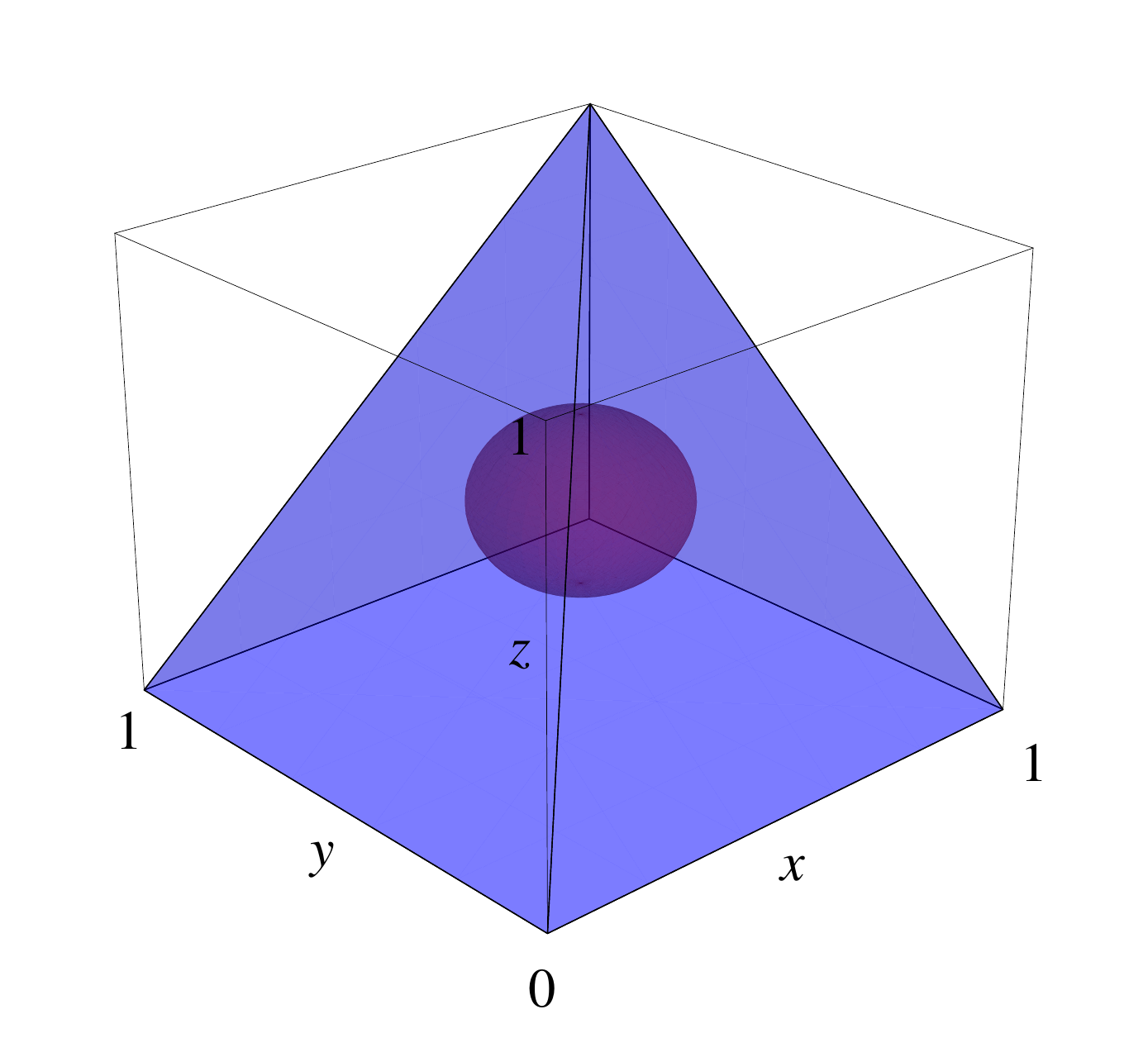}
\hfill
\includegraphics[width=7cm]{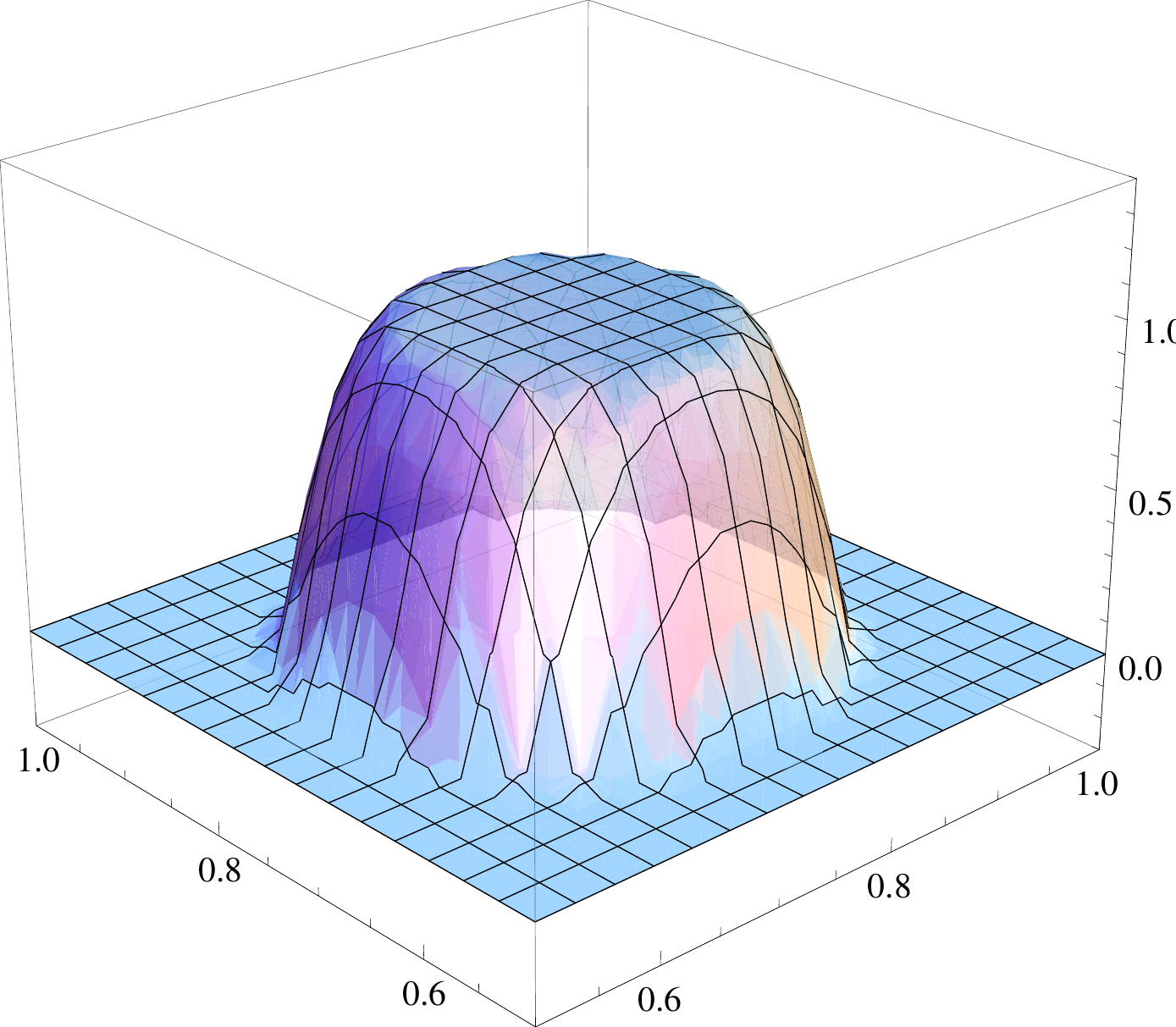}
\caption{
The function $f^A$ is equal to $1$ inside the darker sphere and close to zero elsewhere (left panel);
the graph cut of $f^A$ for $z=\frac{1}{4}$ (right panel).}\label{fig:mollifcharfunc}
\end{figure}

Figure \ref{fig:mollifcharfunc} shows this
function, denoted by $f^A$, for the parameters $\alpha=\frac{1}{10}$, $\beta=\frac{1}{5}$,
$(x_0,y_0,z_0)=(\frac{3}{4},\frac{3}{4},\frac{1}{4})$.
The left picture
shows the sphere where the value of the function $f^A$ is equal to 1;
the right picture shows graph cut for $z=\frac{1}{4}$.

\begin{figure}[htp]
\includegraphics[width=5cm]{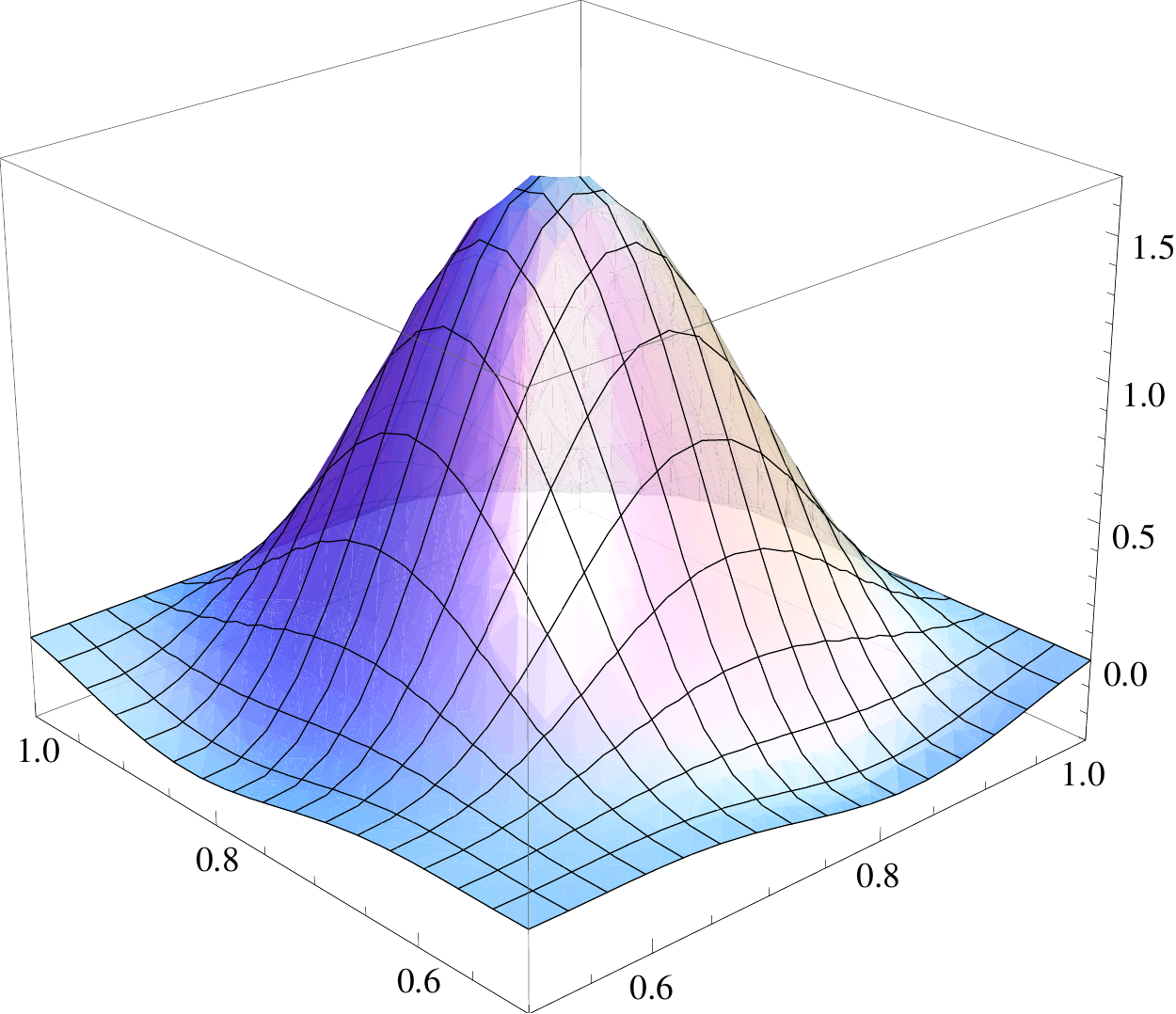}
\hfill
\includegraphics[width=5cm]{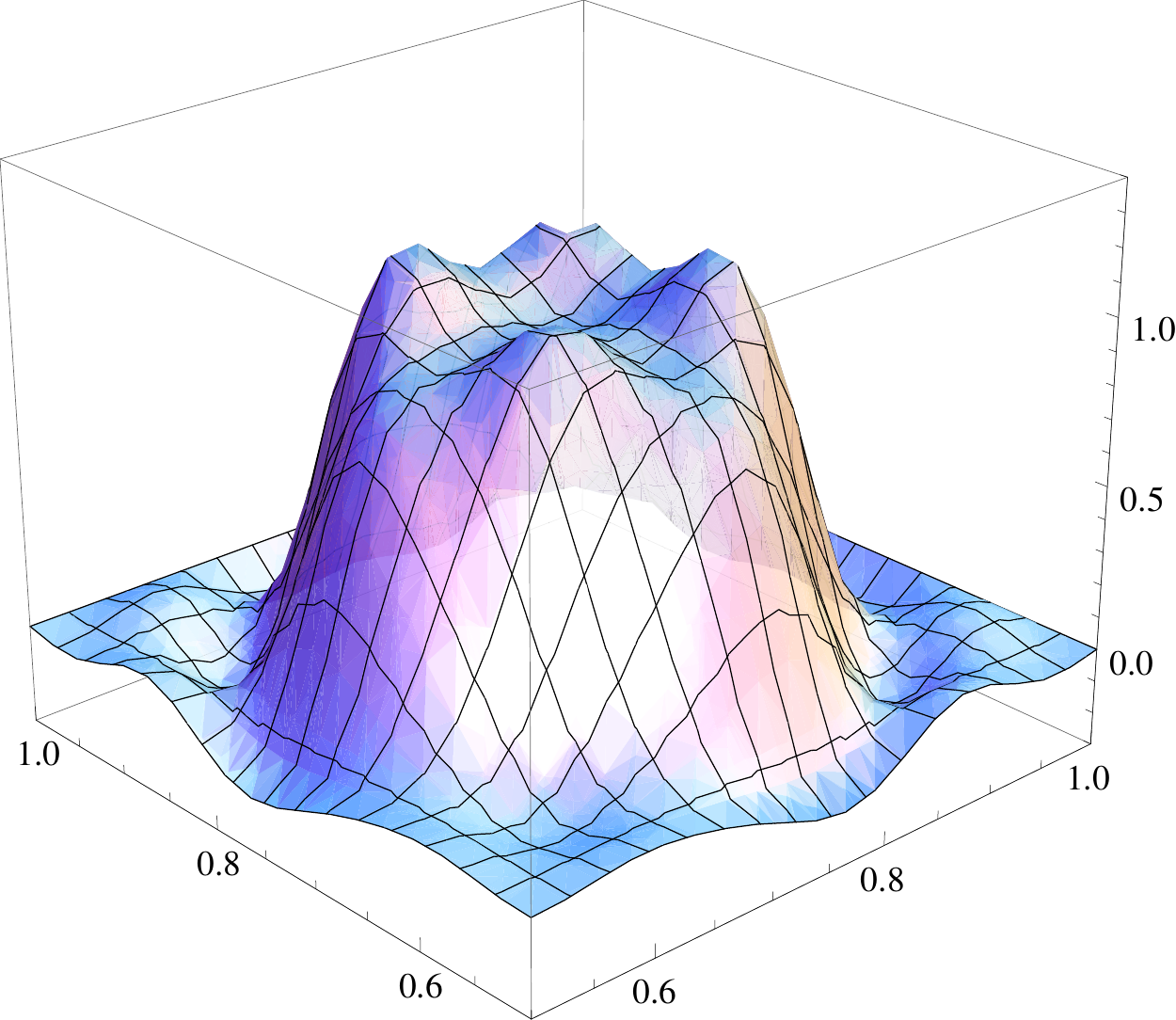}
\hfill
\includegraphics[width=5cm]{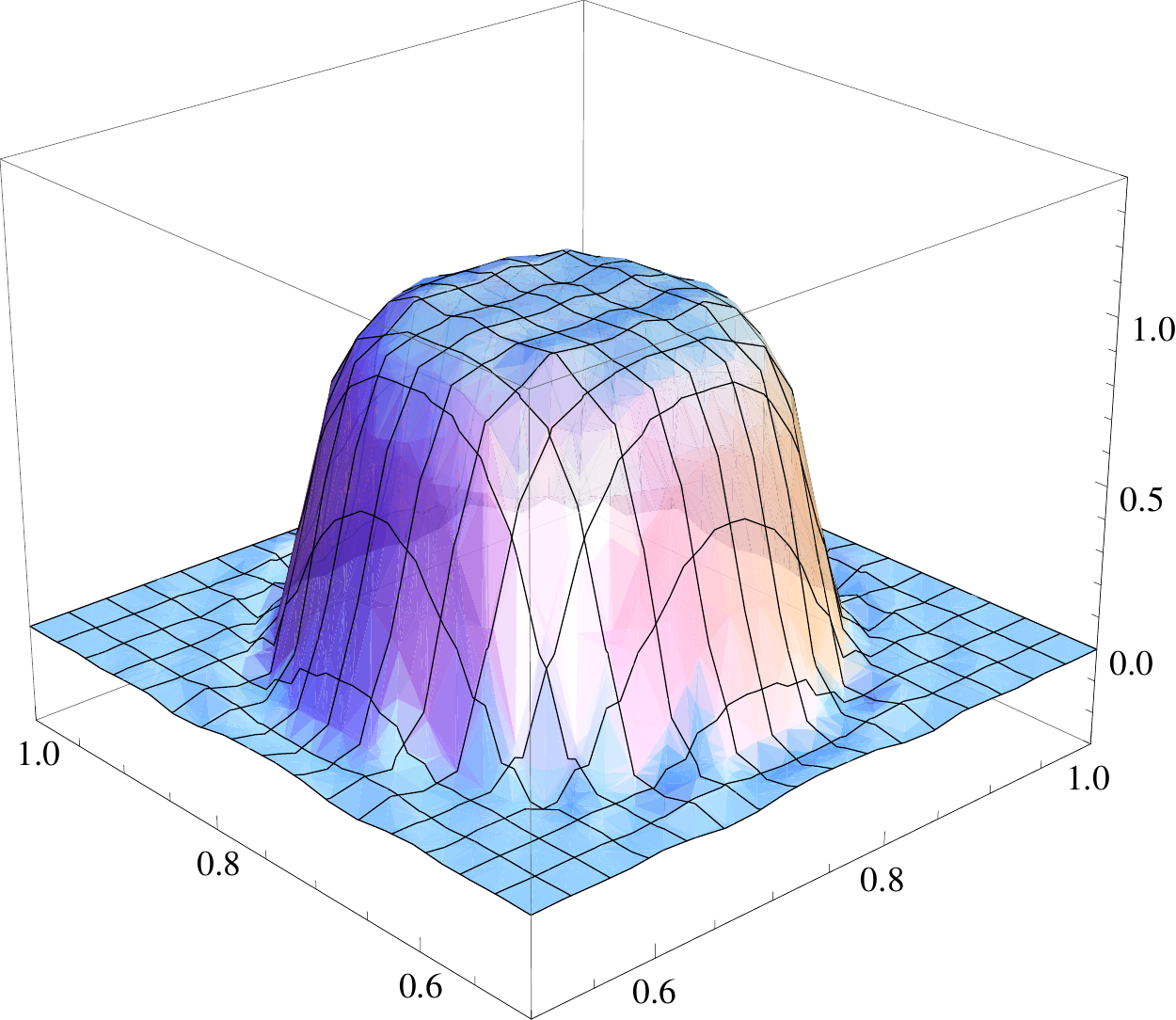}
\caption{The graph cuts ($z=\frac{1}{4}$) of the alternating interpolations $\psi^A_{N}$ of the function $f^A$
for the grid density parameters $N=7,\,15,\,31$.}
\label{fig:interpolations}
\end{figure}

Alternating interpolations of the function $f^A$, sampled on the grids $L_{0,1/2,N,1}$, were calculated for the cases $N=7,\,15,\,31,\,61,\, 121 $. Integral error estimates
 of the form
$
\int_{F(A_3^{\mathrm{aff}})} |f^A-\psi^A_N|^2
$
are listed in Table~\ref{tab:errors}. The resulting interpolating polynomials $\psi^A_N$ for $N=7,\,15,\,31$ are
plotted in Figure~\ref{fig:interpolations}.

\begin{table}[htp]
\centering
\begin{tabular}{c c c c}
\hline\hline
\vrule height13pt depth 10pt width0pt
$N$ & $\int_{F(A_3^{\mathrm{aff}})} |f^A-\psi^A_N|^2$ \\
\hline
\vrule height11pt depth 5pt width0pt
$7$ & $266649\times 10^{-8} $  \\
$15$ & $39178\times 10^{-8} $  \\
$31$ & $2388\times 10^{-8} $   \\
$61$ & $413\times 10^{-8} $   \\
$121$ & $70\times 10^{-8} $  \\
\hline   
\end{tabular}
\caption{Integral error estimates of the interpolations $\psi^A_{N}$ of $f^A$.}
\label{tab:errors}
\end{table}

\section{Concluding remarks}

A family of special functions that is a finitely generated ring can generally be transformed into a polynomial ring. The basis of the ring is taken as the set of variables of the polynomials. Then decomposition of products of the ring of functions allows one to build recursively the polynomials.

In our case that would be the polynomials in $3$ real variables chosen as the lowest alternating functions. Discretization of the functions then provides discrete version of the polynomials. Some information of the discretized $E-$functions of $G_2$ and $C_2$ is already available \cite{MotP,Sz}.  However, probably a logical preference should be given to description of discretized two and three variable $E$-polynomials before the alternating polynomials. That is not found in the literature so far. A comparison of the $E-$ and alternating polynomials should be useful.

Interpolation of discrete data by means of discrete Fourier expansions is an important practical problem for some applications \cite{Stoer, Davis}. Relatively simple possibilities, shown in the paper, deserve to be further analyzed/optimized.

The connection between the (anti)symmetric exponential functions \cite{KPexp} and the (anti)symmetric trigonometric functions \cite{KPtrig} was described in detail for the two dimensional case in \cite{HP,HMP}. An analogous connection is to be expected for the three-dimensional alternating functions and the alternating trigonometric functions \cite{KPalt}.   

In our opinion timely and valuable would be a review and comparison of available  systems of orthogonal functions in two and three variables. Numerous systems are available. For example, there are seven semisimple Lie groups of rank 3. Each has at least two systems of orthogonal functions, in fact majority of them admits four such systems of functions. In addition one should add all the $E-$functions and alternating functions.

\section*{Acknowledgments}
We gratefully acknowledge the support of this work by the Natural Sciences and Engineering Research Council of Canada and by the Doppler Institute of the Czech Technical University in Prague. JH and SP are grateful for the hospitality extended to them at the Centre de recherches math\'ematiques, Universit\'e de Montr\'eal, where most of the work was done. SP acknowledges support by GA Czech Republic (project P201/10/1509). JH also acknowledges support by the Ministry
of Education of Czech Republic (project MSM6840770039). JP expresses his gratitude for the hospitality of the Doppler Institute, where the work was finished.


\end{document}